# Optically Detected Magnetic Resonance of Nitrogen-Vacancy Centers in Diamond Using Two-photon Excitation

Lam T. Nguyen* and Khanh Kieu

*Wyant College of Optical Science, University of Arizona, 1630 E. University Blvd., Tucson, Arizona 85721, USA*

*Corresponding author: ltn99@arizona.edu*

**We demonstrate the use of two-photon excitation for observing the ground state optically detected magnetic resonance (ODMR) of nitrogen-vacancy centers in diamonds at room temperature. An ultrafast femtosecond laser at 1040 nm was used for excitation, while fluorescence signal read out was achieved through a combination of a PMT and a lock-in amplifier. The imaging capability of two-photon excitation fluorescence (2PEF) was utilized to map the distribution of NV centers in a bulk diamond and micro-sized diamonds. For the first time, ODMR traces of the nitrogen-vacancy center are observed with two-photon excitation, providing a promising tool for fast 3D quantum sensing and imaging.**

Diamond is a fascinating crystal of high value in both the commercial market and research owing to its one-of-a-kind physical and quantum properties. A pure defect-free diamond crystal, made of pure carbon atoms arranged in a cubic crystal lattice, appears transparent due to the large bandgap energy (5.5 eV). Atomic impurities introduced to the lattice would alter the optical absorption and emission properties of the diamond, leading to a variety of color centers [1]. Among these centers, nitrogen is the most common impurity, yet the most interesting when accompanied by a single vacancy in the lattice, resulting in a nitrogen-vacancy (NV) center.

The NV center is a point defect in the diamond lattice produced from a substitution of two carbon atoms with a nitrogen atom and a lattice vacancy. There can be different charge states in the NV center based on the number of extra electrons in the system, namely the neutral $NV^0$ center (with no extra electron) and the negative $NV^-$ center (with one extra electron at the vacancy cite). The main interest concerns the $NV^-$ center, which we will refer to as the NV center herein. Thus far, the optical and quantum properties of NV center have been extensively investigated [2-4]. In brief, the ground state of the NV center ($^3A_2$) contains triplet spin states, $m_s = 0$ and $m_s = \pm 1$, based on the spin alignment of the electrons. Once the electron is excited to the excited state ($^3E$), some thermalizations occur before relaxation to the ground state, leading to bright red emission at 637 nm zero phonon line (ZPL) accompanied by a broad phonon sideband ($^3E \rightarrow {}^3A_2$ transition). The population of the $m_s = 0$ ground state can be transferred to the $m_s = \pm 1$ state by applying a resonant microwave (MW) oscillating at around 2.87 GHz. As the excited $m_s = \pm 1$ state decay, they preferably path to an intermediate singlet state ($^3E \rightarrow {}^1A_1$ transition) which is a non-radiative transition, resulting in a decrease in the emitting red fluorescence. This process sets the basis for optically detected magnetic resonance (ODMR) in the NV center of diamond. As the spin states can be read out optically with a long coherent time, NV centers have been continually explored and found valuable applications across various fields, including magnetic sensing [5-6], thermal sensing [7,8], rotation sensing [9], quantum information processing [10-12], EV battery monitoring [13], or memory storage [14].

Until now, one-photon excitation of the NV centers using 515-637 nm light is the primary scheme to harness the magnetic resonance of the NV center [15-17]. While the luminescence can be collected relatively easily and quickly, the NV centers are excited throughout the sample lacking localization. Background fluorescence from other defects and optical aberrations at depth can additionally reduce signal-to-noise ratio, complicating quantitative analysis or degrading of spin properties. That is why specially grown diamonds with a thin near-surface layer of NV centers have to be developed for many applications [18,19]. Confocal microscopy has been extensively employed in ODMR studies of nitrogen-vacancy (NV) centers in diamond [20,21]. The technique allows for imaging with high spatial resolution, capable of resolving a single NV center. However, it can exhibit inherent limitations for imaging deep within bulk diamonds or beneath an interface, where optical aberrations, scattering, and reduced excitation and collection efficiency can significantly degrade spatial resolution and signal fidelity. In particular, the limited penetration depth of green light and the shallow focal volume of confocal microscopy restrict measurements in highly scattering or bulk samples. As a possible alternative, multi-photon microscopy (MPM) is a nonlinear imaging technique known for advantages such as high spatial resolution, high penetration depth, minimal photodamage, and thin intrinsic optical sectioning, leading to 3D mapping capability [22]. Through a two-photon excitation (2PE) process, a femtosecond laser at 1040 nm can provide the photon energy equivalent to that of 520 nm light, suitable to excite the NV centers (Fig. 1(a)).

Previously, it has been shown that near-IR (NIR) excitation of NV centers can be achieved through a 2PE scheme [23,24] and can be applied to obtain high spatial resolution imaging of fluorescent nanodiamonds [25]. However, no ODMR measurement with this NIR excitation scheme has been reported to date. Unlike conventional one-photon photoluminescence (PL), the multi-photon excitation volume is inherently localized. This leads to the 3D mapping capability of MPM. Our previous works have demonstrated high resolution imaging performance at substantial penetration depth within bulk samples, fully covering the mm-scale crystals including diamonds [26,27]. Utilizing the inbuilt optical sectioning capability of MPM, the laser focus can be parked at a single location inside a bulk diamond for ODMR measurement. As a result, the NV centers (both in ensembles and single) distributed throughout the diamond can be imaged in 3D at high speed. In addition, as the fluorescent emissions of most defects are in the visible range, by using an excitation source in the NIR, overlap between excitation and emission light is of minimal concern. The availability of the inherent MW signal due to the regular repetition rate of the excitation femtosecond (fs) pulse train may also add an interesting dimension to NV centers research.

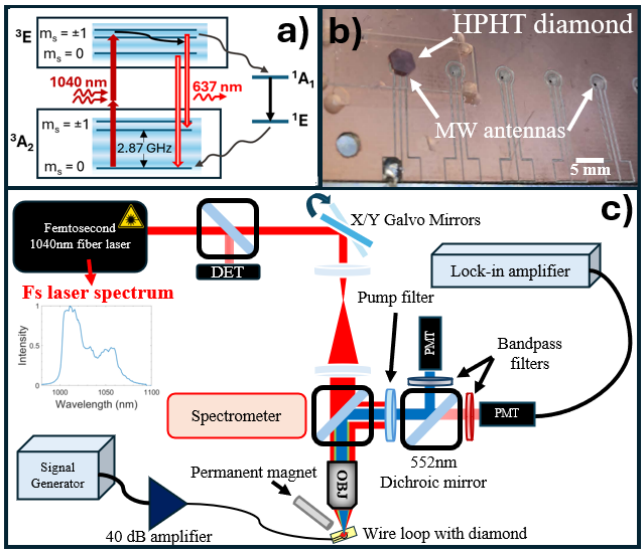

**Fig. 1**. (a) Electronic energy diagram of an NV center when excited with a 1040nm femtosecond laser through a two-photon absorption process. (b) MW delivery assembly with a simple printed circuit board (PCB). The diamond sample is placed in the middle of the hole and secured using a microscope slide epoxied onto the PCB. Imaging is done through the hole as the sample is positioned upside-down. (c) Schematic layout of the apparatus, consisting of the multiphoton microscope with lock-in detection. A home-built 1040 nm Ytterbium fs fiber laser is used to excite the NV-center. The output spectrum of the laser is shown in the inset.

In this Letter, we demonstrate the capability of two-photon excitation fluorescence (2PEF) for diamond imaging, emission spectral analysis, and ODMR observation at room temperature. Additionally, we also imaged micro-sized diamonds with multiple nonlinear channels, empirically showing that these diamonds can exhibit large variations in the concentration of $NV^0$, $NV^-$, and other defect centers. These variations indicate that NV centers are not uniformly distributed in bulk and microdiamonds. As a result, our tool can help identify areas of good NV centers concentration for ODMR measurements.

The experimental setup comprises the multiphoton microscope, microwave delivery, and lock-in detection for fluorescent signal read out (Fig. 1(c)). The MPM system is built in-house with the design and construction detailed elsewhere [28]. The Yb-based fs laser source is similar to the design documented in [29]. Briefly, the 1040 nm laser provides ~50 fs pulses at 7.01 MHz repetition rate with about 60 mW of average power, corresponding to ~8.5 nJ pulse energy. Due to optical losses in the beam path, the pulse energy at the sample plane is ~ 3.5 nJ, which can be reduced further using ND filters. About 1% of the laser power is separated to generate the reference signal for the lock-in amplifier (UHFLI-600). Next, a pair of XY galvo mirrors raster-scans the beam before entering a telescopic beam expander. The laser beam diameter is enlarged 4X to fill the back aperture of the objective (Nikon 20X 0.75 NA), which provides an experimental two-photon lateral and axial resolution of 0.57 µm and 3.95 µm, respectively [30]. An epi-detection scheme is utilized to collect back-propagated nonlinear signals. A rotatable dichroic mirror allows simple switching between the imaging and spectrometer setup. Additional short-pass and band-pass filters are employed to block the residual pump light as well as further separate each nonlinear process into their respective PMT. In this excitation scheme, the emission signal of the NV center corresponds to the 2PEF detection channel, which collects signal ranging from 630 – 950 nm. Other imaging channels such as three-photon excitation fluorescence (3PEF), or second- and third-harmonic generation (SHG and THG), are separated with their respective bandpass filters. The signal readout from each PMT is coupled into the lock-in amplifier (5th order filter and time constant of 30 ms) to monitor the change in fluorescent intensity in ODMR measurements shown below. The time constant is reduced to 2 µs in fast imaging mode.

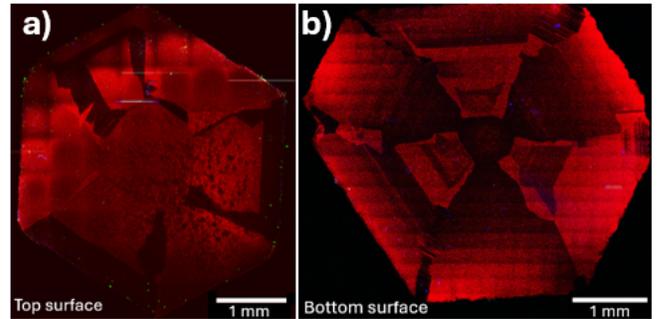

**Fig. 2.** Stitched MPM images of both the (a) top and (b) bottom surface of the red plated HPHT diamond. The hexagonal pattern of the fluorescence signal shows potential effect of the pressure plates during growth and after electron bombardment process. (Green = SHG, Blue = THG, Red = 2PEF, Cyan = 3PEF).

The first sample we imaged is a relatively large (~4x4 mm) red plated single crystal diamond (Adamas nano 1bDPNVB-Macle). It was synthesized by the high pressure and high temperature (HPHT) technique with <111> crystal orientation and ~1.5 ppm of NV concentration. Full surface multi-color multi-photon mosaic, covering both sides of the plated diamond, was conducted to first survey the distribution of NV centers (Fig. 2). We could not image deep below the surface of this sample due to the poor surface roughness (~ 500 nm) of the sample. The full diamond was imaged with a 20X 0.75 NA objective (Nikon) using tiling technique: each individual raster scan (~317 µm FOV) was stitched together to cover the full diamond. False colors and minor image adjustment (brightness and contrast) were added to each nonlinear channel for viewing purposes (red = 2PEF, cyan = 3PEF, blue = THG, green = SHG). As shown by the dominant red color, the majority of the emission contribution originates from the NV centers, which is further validated by the experimental emission spectrum shown in Fig. 3(a). We confirm the nonlinear nature of the optical fluorescence signal by a quadratic power dependence curve for the two-photon process (inset, Fig3(a)). The imaging results reveal interesting zoning artifacts and patterns on both top and bottom surfaces. These artifacts could have originated from the HPHT growth process, the high energy electron beam irradiation and thermal annealing to create NV centers.

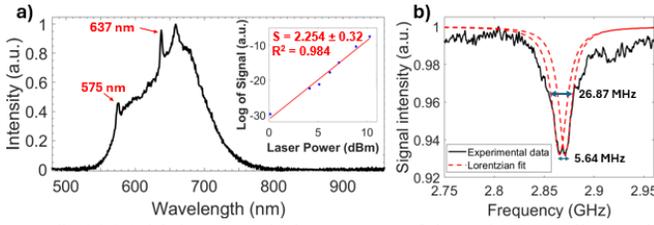

**Fig. 3.** (a) Multiphoton emission spectra of the red HPHT diamond. The ZPL emission from the $NV^0$ centers and the $NV^-$ centers are distinctively observed at 575 nm and 637 nm, respectively. The inset validates the nonlinear nature of the emission signal, with the power dependence slope closely following a quadratic relationship (Slope = 2.25 ± 0.32). (b) ODMR trace in absence of an external magnetic field, fitted with two Lorentzian curves.

To observe the zero-field splitting magnetic resonance, we monitored the demodulated signal from the lock-in amplifier as the MW frequency was gradually swept between 2.75-2.96 GHz over a period of 40 seconds. The RF signal originated from a signal generator (Anritsu 68369A/NV) producing up to +5 dBm of RF power. This microwave was then put through a ~40 dB gain microwave amplifier (ZHL-4240+), resulting in ~30 dBm average power at the MW loop without factoring in the MW insertion and reflection losses. To deliver the sweeping RF signal to the diamond, we constructed a MW antenna by using a trimmed BNC cable, connected via soldering to a 0.8 mm-thick copper printed circuit board (PCB). The MW loop on the PCB was created with a CNC machine. The center hole diameter is ~1.77 mm where the diamond was placed atop in direct contact with the copper MW loop (Fig. 1(b)). The diamond was secured by attaching a microscope slide on the back side, then adhered onto the PCB with a quick cure epoxy. Imaging was conducted through the hole as the working distance of the objective is longer than the thickness of the PCB. With this arrangement, the ODMR signal was collected from the "top surface" of the HPHT diamond as denoted in Fig. 2(a).

The ODMR trace can be seen in Fig. 3(b), fitted with two Lorentzian fitting curves (the average fs laser power on the diamond was ~4.2 mW). The FWHM linewidth of the fluorescence dip was measured to be ~26.87 MHz, separated by the hyperfine splitting of the $m_s = \pm 1$ states (~5.64 MHz). This naturally occurring splitting is due to the strain in the crystal lattice, affecting the magnetic spin coupling between the defect and nearby spins and lifting the degeneracy of the electronic sublevels [3]. The reduction of the fluorescent intensity was ~7%. The traces were stable and repeatable over many MW sweeping cycles indicating the stability of NV centers under our femtosecond laser interaction.

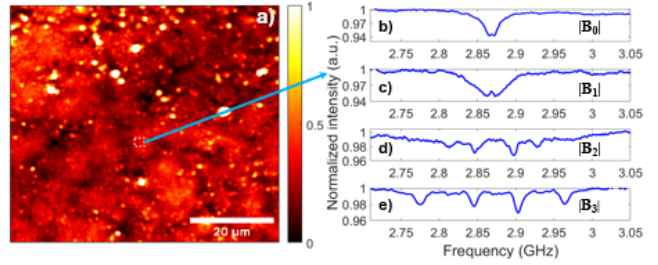

**Fig. 4.** (a) A zoomed-in region on the top surface of the red HPHT diamond, showing the lack of emission uniformity. (b-e) ODMR traces collected from the same location (specifically from the white dotted box in (a)) with increasing magnetic field intensity. The magnet was positioned at an arbitrary orientation with respect to the crystal coordinate system. The resonant frequencies of the fluorescence dips shift due to Zeeman splitting when the static magnetic field is increased near the vicinity of the diamond.

We also explored the noticeable effect on the ODMR spectra due to Zeeman splitting of the NV center when a static magnet was brought into the vicinity of the bulk diamond. The signal was collected from a small focal volume (~13 μm³) of the bulk HPHT diamond, denoted by the white dotted box in Fig. 4(a). Fig. 4(b-e) shows the ODMR spectra as the strength of the magnetic field was increased in the order of $|B_0|<|B_1|<|B_2|<|B_3|$ with $|B_0|$ being the absence of the magnetic field (i.e. $|B_0| \approx 0$). Prior to the introduction of the magnetic field, the ODMR spectrum was as expected, two closely spaced fluorescent dips of ~6%, centered at 2.87 GHz and separated by the hyperfine splitting of ~5 MHz (Fig. 4(b)). A drastic change in the spectra, both in the resonant frequency and contrast of the fluorescent dip, was observed as the magnet was brough closer to the diamond. The four-dip pattern indicates the magnetic field projected two unequal and non-zero components into the crystal coordinate system with the last two components being zero [31]. We also observed other ODMR splitting patterns (e.g. 6 dips or 8 dips) by repositioning the orientation of the magnet (not shown here).

Next, we also imaged and collected ODMR spectra of individual micro-sized diamonds. The 15 μm diamonds (Adamas nano MDNV15umHi30mg) are in powdery form with ~3.5 ppm concentration of NV centers, as specified by the vendor. They were deposited on a thin microscope coverslip and placed on top of the MW antenna in similar arrangement to the plated diamond. We image the microdiamonds directly on top of the hole here, not through it as in the case of the plated diamond above. The distance from the microdiamonds to the MW loop is about 130 μm (the thickness of the microscope coverslip). Multi-color imaging result can be observed in Fig. 5(a). Interestingly, the distribution of the NV centers across different microdiamonds is clearly inhomogeneous, characterized by the varying 2PEF intensity (false colored in red). Further exploring the non-uniformity through spectral measurements of a couple of diamonds in the FOV (Fig. 5(b)), some can be seen to possess significantly higher concentration of $NV^0$ centers (ZPL at 575 nm). In others, $NV^-$ centers (ZPL at 637 nm) are the main contributors to the 2PEF signal. Apart from the NV centers, we also observed other nonlinear signals such as the harmonic signals (SHG and THG) or other unidentified 3PEF signals (with well-defined peaks at 388 nm, 398 nm, and 408 nm), reaffirming that MPM can effectively be used as a screening tool for different defect centers in diamond. We also collected ODMR spectra from both populations of microdiamonds. As expected, the characteristic dip in fluorescence was observed exclusively in those with higher concentration of $NV^-$ centers (e.g., microdiamond 2), while those lacking the characteristic ZPL of the $NV^-$ center (e.g., microdiamond 1) exhibited no measurable variation in fluorescence intensity (not shown here). Fig. 5(c) provides ODMR result from microdiamond 2, fitted with 2 Lorentzian functions, indicating a 2.5 % change in fluorescence intensity and a hyperfine splitting of ~8 MHz (E = 4 MHz).

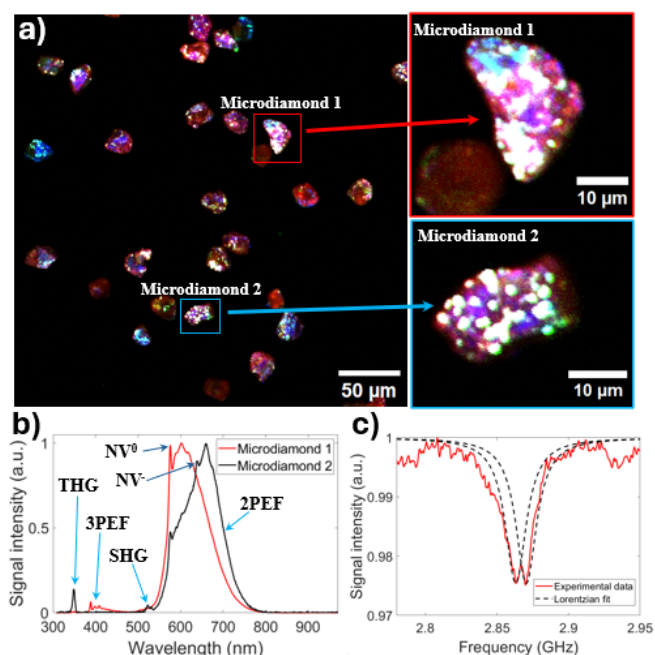

**Fig. 5.** Imaging, spectral measurement, and ODMR spectra of micro-diamonds. (a) Multi-color image of micro-sized diamonds (Green = SHG, Blue = THG, Red = 2PEF, Cyan = 3PEF). The white-colored portions indicate an overlap of three or more nonlinear signals. (b) Representative spectra of 2 different diamonds in the FOV, showing one with mostly $NV^0$ centers while the other clearly has higher concentration of $NV^-$ centers. (c) ODMR trace of micro diamond 2.

In conclusion, we have demonstrated that two-photon excitation of the NV centers using a 1040 nm fs source can be efficiently utilized both for imaging and obtaining localized ODMR spectra from different forms of diamond. By employing this excitation scheme, 2PE can enable efficient and localized excitation of NV centers, thereby facilitating high resolution and fast mapping of diamond samples. Clear ODMR traces were observed, showing zero-field splitting, hyperfine splitting and Zeeman splitting. These attributes make 2PEF a promising approach for high-resolution, 3D quantum sensing, imaging and other NV-based applications.

**Disclosures.** The authors declare no conflicts of interest.

**Data availability.** Data underlying the results presented in this Letter are not publicly available at this time but may be obtained from the authors upon reasonable request.